\documentclass[twocolumn]{article}
\usepackage[margin=0.6in,columnsep=0.15in]{geometry}
\usepackage[utf8]{inputenc}

\usepackage[usenames,dvipsnames]{xcolor}
\usepackage[colorlinks,citecolor=Gray,urlcolor=Gray,linkcolor=Blue]{hyperref}

\usepackage{graphicx}
\usepackage{mathtools}
\usepackage{booktabs}
\usepackage{tabularx}
\usepackage{tikz}
\usepackage{color}
\usepackage{siunitx}
\usepackage[lcgreekalpha,upint]{stix}

\usepackage[tiny]{titlesec}
\usepackage[inline]{enumitem}

\usepackage[normal]{caption}
\DeclareCaptionLabelSeparator{pipe}{ $|$ }
\usepackage{authblk}

\usepackage[colon]{natbib}
\usepackage{doi}

\title{\Large Density-functional model for van der Waals interactions:\\Unifying many-body atomic approaches with nonlocal functionals}

\author[1,2,*]{Jan Hermann}
\author[1,*]{Alexandre Tkatchenko}
\affil[1]{Department of Physics and Materials Science, University of Luxembourg,
L-1511 Luxembourg City, Luxembourg}
\affil[2]{FU Berlin, Department of Mathematics and Computer Science, Arnimallee 6,
14195 Berlin, Germany}

\date{}

\makeatletter
\def\blfootnote{\xdef\@thefnmark{}\@footnotetext}
\makeatother

\begin{document}

\twocolumn[{%
  \maketitle
  \vspace{-3em}
  \begin{center}
  \begin{minipage}{0.85\linewidth}
    \small
    \paragraph{Abstract}
    Noncovalent van der Waals (vdW) interactions are responsible for a wide
    range of phenomena in matter. Popular density-functional methods that
    treat vdW interactions use disparate physical models for these intricate
    forces, and as a result the applicability of these methods is often
    restricted to a subset of relevant molecules and materials. Aiming
    towards a general-purpose density functional model of vdW interactions,
    here we unify two complementary approaches: nonlocal vdW functionals for
    polarization and interatomic methods for many-body interactions. The
    developed nonlocal many-body dispersion method (MBD-NL) increases the
    accuracy and efficiency of existing vdW functionals and is shown to be
    broadly applicable to molecules, soft and hard materials including ionic
    and metallic compounds, as well as organic/inorganic interfaces.
  \end{minipage}
  \end{center}
  \vspace{1em}
}]

\blfootnote{$^*$Emails: science@jan.hermann.name, alexandre.tkatchenko@uni.lu}%

\newcommand{\citein}{\citealp}
\newcommand{\SMref}{Appendix}
\newcommand{\thiswork}{work}



Van der Waals (vdW) interactions originate from nonlocal correlations in the quantum motion of electrons and give rise to a wide spectrum of physical phenomena from attraction between two atoms \citep{LondonZP30} to the macroscopic Casimir effect \citep{JaffePRD05}.
As a result, vdW interactions are one of the prime targets in material modeling, which has led to a plethora of approaches that either treat vdW forces in the same way as the rest of electron correlation, or model them with effective potentials \citep{KlimesJCP12,GrimmeCR16,HermannCR17}.
They include quantum Monte--Carlo (QMC) \citep{AmbrosettiJPCL14}, coupled cluster methods \citep{YangS14}, random-phase approximation \citep{LuPRL09}, nonlocal density functionals \citep{DionPRL04,VydrovPRL09}, and coarse-grained approaches ranging from pairwise \citep{GrimmeJCP10,BeckeJCP07,TkatchenkoPRL09} to many-body models \citep{TkatchenkoPRL12,SilvestrelliJCP13,CaldeweyherJCP19}.

From a theoretical perspective, this status quo is undesirable, because different models offer often disparate pictures of the nature of vdW forces, leading to incoherent understanding of vdW interactions in molecules and materials.
From a practical perspective, the three main characteristics of a method are its generality, accuracy, and computational efficiency, and so far, no single method has satisfied all three requirements while being applicable to all relevant types of matter.
For instance, QMC and coupled cluster are limited by computational efficiency, pairwise approaches and two-point vdW functionals lack in accuracy for nanostructured and supramolecular compounds, and atomic models have qualitative problems with ionic and hybrid metal-organic systems.

In this \thiswork, we present a unified density-functional model of vdW interactions that couples polarizability density functionals and atomic models, inheriting broad applicability of the former and excellent accuracy of the latter.
We integrate the polarizability functional of \citet{VydrovPRA10} (VV), normalization to exact free-atom vdW parameters of the Tkatchenko--Scheffler (TS) model \citep{TkatchenkoPRL09}, normalization to jellium via a zero-gradient limit from the VV10 functional \citep{VydrovJCP10a}, and the Hamiltonian for the dispersion energy of the many-body dispersion (MBD) model \citep{TkatchenkoJCP13}.
Compared to the range-separated self-consistently screened (rsSCS) variant of MBD \citep{AmbrosettiJCP14}, the VV polarizability functional enables consistent treatment of ionic compounds, normalization to the free-atom reference balances the accuracy of the VV polarizability across the periodic table, and normalization to jellium enables effective modeling of metals and their surfaces \citep{RuizPRL12}.
The new model involves a similar level of empiricism as MBD@rsSCS---we remove the tabulated vdW radii and short-range screening, while introducing a mechanism to avoid double counting of electron correlation in near-uniform density regions.
We demonstrate on a series of benchmark calculations that the new model enables for the first time a consistent treatment of vdW interactions in molecular, covalent, ionic, metallic, and hybrid metal-organic systems.

Some of the problems of MBD@rsSCS have been previously treated by \citet{GouldJCTC16a}.
Their fractionally ionic variant of MBD@rsSCS uses iterative Hirshfeld partitioning in combination with a piecewise linear dependence of atomic polarizability on charge, together with a rescaling scheme for the diverging MBD Hamiltonian in highly polarizable systems or with dipole smearing \citep{KimJACS20}.
Our approach is instead based on a general polarizability functional.




Atomic models, such as MBD, require an atomic response model in the form of static polarizabilities $\alpha_{0,i}\equiv\alpha_i(0)$ and $C_{6,ii}$ coefficients.
In the new model, dubbed MBD-NL, we parametrize the response of atoms by coarse-graining the VV polarizability density to atomic fragments \citep{HirshfeldTCA77,SatoJCP09,SatoJCP10}.
The VV polarizability functional is a semilocal functional of the electron density $n(\mathbf r)$, which models the local dynamic polarizability density \citep{VydrovPRA10},
\begin{equation}
  \alpha^\text{VV}[n](\mathbf r,\mathrm iu)
    =\frac{n(\mathbf r)}{\frac{4\pi}3n(\mathbf r)
    +C\frac{{|\boldsymbol\nabla n(\mathbf r)|}^4}{n{(\mathbf r)}^4}+u^2}
  \label{eq:vv-functional}
\end{equation}
where $\mathrm iu$ is imaginary frequency and $C$ is an empirical parameter.
The atomic dynamic polarizabilities are obtained by partitioning the polarizability density with Hirshfeld weights $w_i^\text{H}(\mathbf r)=n_i^\text{free}(\mathbf r)/\sum_j n_j^\text{free}(\mathbf r)$,
\begin{equation}
  \alpha_i^\text{VV}(\mathrm iu)
    =\int\mathrm d\mathbf r\,w_i^\text{H}(\mathbf r)\alpha^\text{VV}[n](\mathbf r,\mathrm iu)
\end{equation}
The $C_6$ coefficients are then calculated directly from $\alpha_i(\mathrm iu)$ via the Casimir--Polder formula \citep{McLachlanPRSLA63},
\begin{equation}
  C_{6,ii}^\text{VV}=\frac3\pi\int_0^\infty\mathrm du\,\alpha_i^\text{VV}{(\mathrm iu)}^2
\end{equation}
Unlike approaches that use Hirshfeld fragments to define atomic volumes, MBD-NL is independent of the choice of a particular atomic partitioning, because this influences only local redistribution of the polarizability between atoms, conserving the total polarizability.
Our approach is also different from that of \citet{SilvestrelliPRL08}, where the electron density is coarse-grained first and a polarizability functional is evaluated over the fragment densities.

Already this bare combination of MBD and the VV polarizability functional substantially improves the description of ionic systems compared to MBD@rsSCS, because the VV functional gives a good estimate of the ionic polarizabilities, unlike the Hirshfeld volume scaling used in MBD@rsSCS\@.
However, this bare combination suffers from two fundamental shortcomings.
First, the polarizability functional is not evenly accurate across the periodic table.
Second, when combined with semilocal density-functional theory (DFT), it suffers from double counting of electron correlation in regions of slowly-varying electron density.
To solve these two challenges, we normalize the atomic VV polarizabilities and $C_6$ coefficients to exact values for free atoms \citep{TkatchenkoPRL09}, and then normalize MBD-NL to give zero vdW energy for jellium by subtracting the portion of the polarizability from slowly-varying electron-density regions.


\begin{figure}[t!]
\centering
\includegraphics{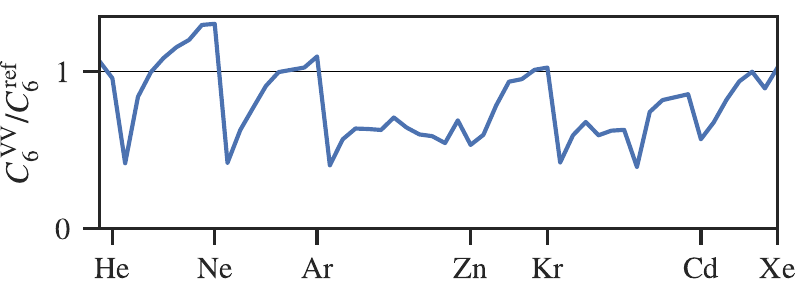}
\caption{%
\textbf{Relative errors in $C_6$ coefficients of free atoms calculated with the VV polarizability functional for the first 54 elements.}
The reference values are taken from the TS method \citep{TkatchenkoPRL09}.
In contrast, the present model is exact by construction (Eq.~\ref{eq:ts-scaling}).
}\label{fig:vv-periodic-table}
\end{figure}

The VV polarizability functional is approximate, which is manifest already for free-atom polarizabilities and $C_6$ coefficients, where accurate reference values are known (Figure~\ref{fig:vv-periodic-table}).
It especially underestimates the vdW parameters of metallic elements.
To mitigate this, we normalize the VV atomic quantities with the ratio of the respective free-atom values obtained from accurate reference calculations and from the VV functional,
\begin{equation}
  \alpha_{0,i}^\text{rVV}
    =\alpha_{0,i}^\mathrm{VV}
    \frac{\alpha_{0,i}^\text{ref,free}}{\alpha_{0,i}^\text{VV,free}},\quad
  C_{6,ii}^\text{rVV}
    =C_{6,ii}^\mathrm{VV}\frac{C_{6,ii}^\text{ref,free}}{C_{6,ii}^\text{VV,free}}
  \label{eq:ts-scaling}
\end{equation}


Many exchange--correlation (XC) functionals are exact for jellium by construction, even though the portion of electron correlation from the nonlocal plasmons is long-ranged and should not be included in semilocal XC functionals.
As a result, most XC functionals describe accurately the electron correlation \emph{within} slowly-varying density regions, such as found in metals, and those cases require no addition of vdW forces.
This is different in most general systems, in which semilocal functionals neglect long-range vdW interactions.
At the same time, these metallic-density regions contribute dominantly to the VV polarizability, and hence to the vdW energy in any vdW model that would use the VV functional directly.
When combined with semilocal DFT, this would result in overpolarization and overbinding of bulk metals, and of adsorbates on metallic surfaces.
To avoid this double counting, the VV10 expression for the vdW energy subtracts the limit of VV10 as the density gradient approaches zero \citep{VydrovJCP10a}, $E_\text{vdW}^\text{VV10}=E^\text{VV10}[n]-\bigl(E^\text{VV10}|_{\boldsymbol\nabla n\rightarrow0}\bigr)[n]$.
Such an approach cannot be used directly in a many-body model such as MBD, because unlike in a pairwise model the many-body vdW energy is not linear in the polarizability.

To ensure the correct zero limit of MBD-NL for uniform densities, we smoothly cut off the contribution of jellium-like regions to the polarizability.
These regions can be distinguished with the combination of two local electron-density descriptors: the local ionization potential $I$ \citep{GutleIJQC99} and the iso-orbital indicator $\chi$ \citep{BeckeJCP90,KummelMP03,SunPRL13},
\begin{equation}
  I[n]=\frac{\tau^\text{W}[n]}n,\qquad
  \chi[n]=\frac{\tau^\text{KS}[n]-\tau^\text{W}[n]}{\tau^\text{unif}[n]}
\end{equation}
where $\tau^\text{KS}[n]=\sum_i|\boldsymbol\nabla\phi_i|^2/2$ is the positive kinetic energy density of occupied orbitals $\phi_i$, which for single-orbital densities reduces to the von Weizsäcker kinetic energy density, $\tau^\text W[n]=|\boldsymbol\nabla n|^2/8n$, and for jellium to $\tau^\mathrm{unif}[n]=3(3\pi^2)^{2/3}n^{5/3}/10$.  
The local ionization potential is a form of a reduced gradient with the units of energy, which attempts to model the electronic gap locally.
The density gradient alone is insufficient to characterize metallic densities.
In particular, both $I\sim0$ and $\chi\sim1$ must be true for density to be metallic, whereas $I\sim0$ and $\chi\sim0$ corresponds to centers of covalent bonds, and $I\sim0$ and $\chi\gg1$ signifies overlap of electron-density tails between noncovalently bound systems.
Since the normalization of VV10 to jellium uses only the density gradient, it partially omits contributions from covalent bonds.
By using also the iso-orbital indicator, we make MBD-NL more precise in this regard.
In practical calculations, the evaluation of the kinetic energy density is the computationally most demanding part of MBD-NL, but this means that its cost is only a fraction of a single self-consistent cycle of a regular meta-GGA KS-DFT calculation.

\begin{figure}[t!]
\centering
\begin{tikzpicture}
\node[below right] at (0,0) {\bfseries a};
\node[below right, inner sep=0pt] at (0,0) {\includegraphics{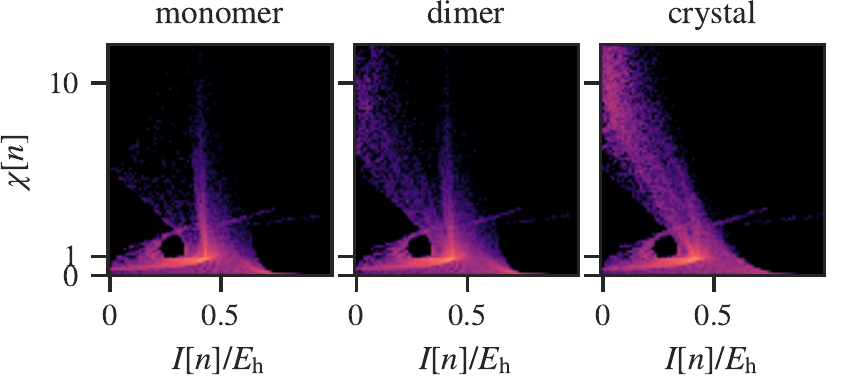}};
\node[below right] at (0,-4.3) {\bfseries b};
\node[below right, inner sep=0pt] at (0,-4.3) {\includegraphics{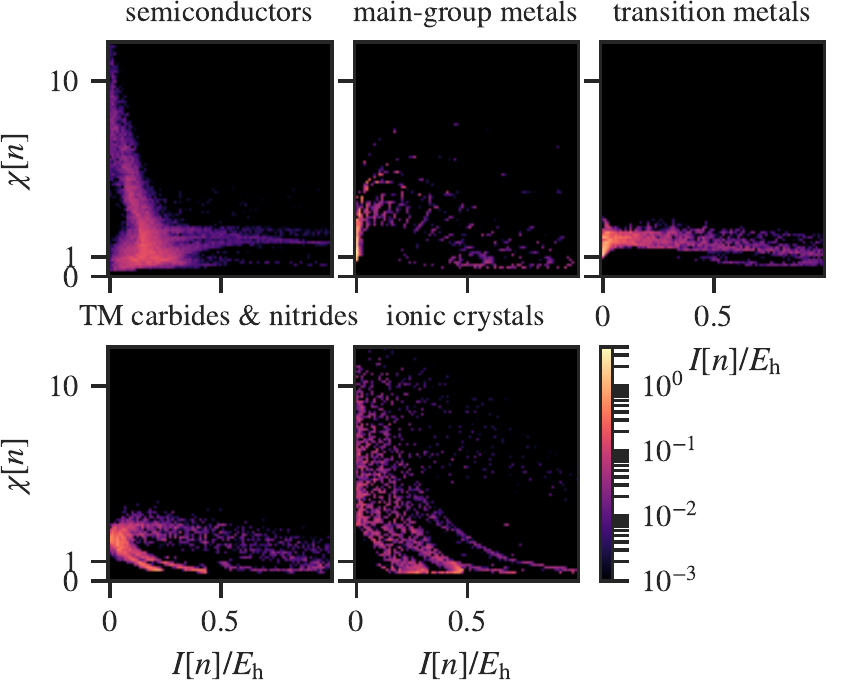}};
\end{tikzpicture}
\caption{\textbf{Polarizability distributions of local ionization potential $I$ and iso-orbital indicator $\chi$.}
The plotted distributions are $\alpha^\text{VV}(s',\chi')=\int\mathrm d\mathbf r\delta(s(\mathbf r)-s')\delta(\chi(\mathbf r)-\chi')\alpha^\text{VV}(\mathbf r)$, such that the total polarizability is $\iint\mathrm ds\mathrm d\chi\,\alpha^\text{VV}(s,\chi)$.
$E_\text{h}$ is one hartree.
(\textbf a) Benzene monomer, dimer, and crystal.
Each distribution is normalized to one benzene molecule.
(\textbf b) 64 simple solids divided to five groups \citep{ZhangNJP18}.
Each distribution is normalized to 62 (a.\,u.), the VV polarizability of a benzene monomer, for a single color scale with \textbf{a}.
}\label{fig:pol-hists}
\end{figure}

Figure~\ref{fig:pol-hists}a presents polarizability density distributions of $I$ and $\chi$ in three benzene compounds and in a set of simple solids \citep{ZhangNJP18}.
In an organic molecule such as benzene (Figure~\ref{fig:pol-hists}a), the vast majority of the polarizability comes from electron density with $I>\SI{5}{\electronvolt}$, with a small part from low-gradient regions with $\chi<1$.
The intermolecular interactions in the benzene dimer and crystal add a significant amount of polarizability in regions with $\chi\gg1$, despite the electron density being low there.
A richer spectrum of patterns is found in simple solids (Figure~\ref{fig:pol-hists}b).
Most similar to the benzene compounds are semiconductors.
In contrast, the polarizability in main-group metals is dominated by jellium-like regions near $(I,\chi)=(0,1)$.
In transition metals, the polarizability is distributed in a wider range of the local gap along the $1<\chi<2$ strip, with a larger part still in the low-gradient regions.
In simple ionic solids, most of the polarizability comes from single-orbital regions ($\chi<1$).

To avoid the double counting of vdW interactions of low-gradient densities, we smoothly cut off their contribution to the polarizability functional,
\begin{equation}
  \alpha^\mathrm{VV'}[n](\mathbf r)=g(I,\chi)\alpha^\text{VV}[n]
\end{equation}
We impose two simple requirements on this cutoff.
First, the density regions with a local gap lower than the work function of conductors should not contribute to the calculated vdW energy, because those are assumed to be covered by a semilocal XC functional.
We chose the cutoff value of 5\,eV, which is around the peak of the work function of elemental metals.
Second, the VV polarizability of simple covalent compounds (exemplified by a benzene molecule) should not be influenced by the cutoff.
The following function $g$ satisfies these two requirements:
\begin{equation}
\begin{gathered}
  g(I,\chi)=1-\frac{1-f\bigl(\chi-3\sqrt{I/E_\text{h}}\bigr)}{1+\exp\bigl(4(I-\SI{5}{\electronvolt})/\SI{1}{\electronvolt}\bigr)}, \\
  f(x)=\exp\bigl(-\theta(x)cx/(1-x)\bigr)\theta(1-x)
\end{gathered}
\end{equation}
Function $g$ consists of a logistic function centered at $\SI{5}{\electronvolt}$ and of function $f$ taken from the SCAN functional \citep{SunPRL15}, where it also interpolates between $\chi=0$ and $\chi=1$ (see \SMref\ for a plot of $g(I,\chi)$).
We find that $c=1/10$ ensures that the effect of the cutoff on the VV polarizability of a benzene molecule is negligible ($<2\%$).
The performance of the resulting model depends only weakly on the precise values of the chosen parameters, as long as the local gap cutoff sufficiently covers the work function of a given conductor.
Nevertheless, a more rigorous formulation of the model in this direction would be desirable.

\begin{figure}[t!]
\centering
\includegraphics{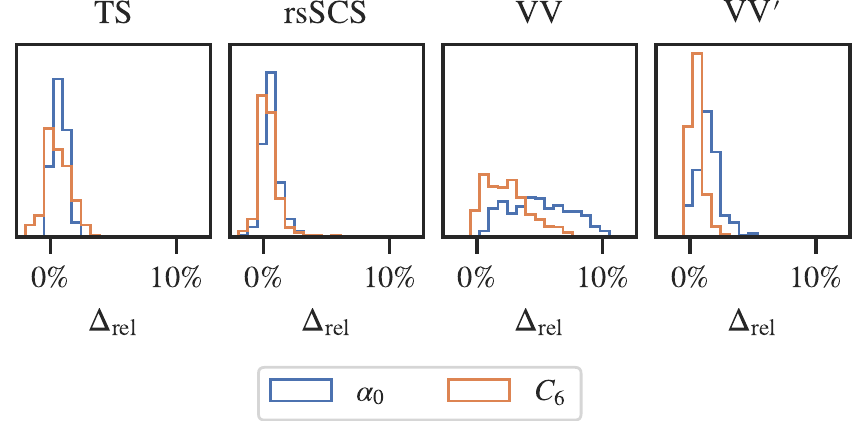}
\caption{\textbf{Distributions of relative changes in atomic static polarizabilities and $C_6$ coefficients from monomers to dimers.}
The distributions are calculated over all atoms from all complexes in the S66 data set \citep{RezacJCTC11}.
}\label{fig:pol-shifts}
\end{figure}

Apart from avoiding the double counting of long-range electron correlation, the cutoff function removes another deficiency of the VV polarizability functional.
When molecules form vdW-bound compounds, the introduction of density-tail overlaps significantly increases the VV polarizability compared to the monomers (Figure~\ref{fig:pol-hists}a).
This effect is an artifact of the VV functional that causes increasingly large vdW-bound systems to be overbound, and cutting off the polarizability of low-gradient regions with $\chi>1$ eliminates this issue without affecting the polarizabilities of isolated monomers (Figure~\ref{fig:pol-shifts}).


Finally, the static polarizabilities and $C_6$ coefficients calculated as described above are directly used in the MBD Hamiltonian to obtain the vdW energy.
This Hamiltonian describes a system of charges in harmonic potentials---Drude oscillators---characterized by their static polarizabilities $\alpha_{0,i}$ and resonance frequencies $\omega_i=4C_{6,ii}/3\alpha_{0,i}^2$, and interacting via a long-range dipole potential $\mathbf T^\mathrm{lr}(\mathbf R)\equiv f_\text{lr}(R)\mathbf T(\mathbf R)$ \citep{LucasP67,TkatchenkoPRL12},
\begin{multline}
  H^\text{MBD}(\{\alpha_{0,i},\omega_i\})
  =\sum_i-\frac12\nabla_{\xi_i}^2+\sum_i\frac12\omega_i^2\xi_i^2 \\
  +\frac12\sum_{ij}\omega_i\omega_j\sqrt{\alpha_{0,i}\alpha_{0,j}}\boldsymbol{\xi}_i\cdot\mathbf T^\mathrm{lr}_{ij}\boldsymbol{\xi}_j
\end{multline}
where $\boldsymbol\xi_i\equiv\sqrt{m_i}\mathbf x_i$ are displacements of the charges weighted with masses $m_i$ (which have no effect on the energy).
The interaction energy of this model system---the vdW energy---is obtained by direct diagonalization yielding a set of coupled oscillation frequencies $\tilde\omega_k$,
\begin{equation}
  E_\text{MBD}=\sum_k^{3N}\frac{\tilde\omega_k}2-\sum_i^N\frac{3\omega_i}2
\end{equation}

In MBD-NL, we use the same long-range coupling $\mathbf T^\text{lr}$ as in the MBD@rsSCS variant \citep{AmbrosettiJCP14},
\begin{equation}
  f_\text{lr}(R_{ij})=1\Big/\Big(1+\mathrm e^{-6{\textstyle(}R_{ij}/\beta(R_i^\text{vdW}+R_j^\text{vdW})-1{\textstyle)}}\Big)  
\end{equation}
but with a simplified definition of the vdW radii.
Rather than tabulated vdW radii, we use the quantum-mechanical formula for vdW radii of free atoms from \citet{FedorovPRL18}, and scale them similarly to the vdW parameters as in~\eqref{eq:ts-scaling},
\begin{equation}
  R_i^\text{vdW}=\tfrac52{(\alpha_{0,i}^\text{ref,free})}^\frac17{\left(\frac{\alpha_i^\mathrm{rVV'}}{\alpha_i^\text{rVV$'$,free}}\right)}^\frac13
\end{equation}
We optimize the damping parameter $\beta$ of $f_\text{lr}(R)$ on the S66 data set \citep{RezacJCTC11}, as was done for MBD@rsSCS, and find the optimal values of 0.81 and 0.83 for the XC functionals PBE \citep{PerdewPRL96} and PBE0 \citep{AdamoJCP99}, respectively, only slightly smaller than the values of 0.83 and 0.85 for MBD@rsSCS\@.


\begin{figure}[t!]
\centering
\begin{tikzpicture}
\node[below right] at (0,0) {\bfseries a};
\node[below right, inner sep=0pt] at (0,0) {\includegraphics{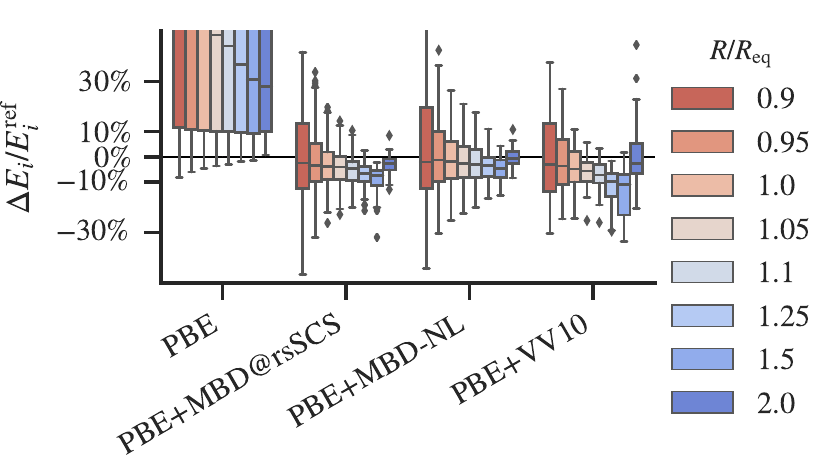}};
\node[below right] at (0,-4.8) {\bfseries b};
\node[below right, inner sep=0pt] at (0,-5.1) {\includegraphics{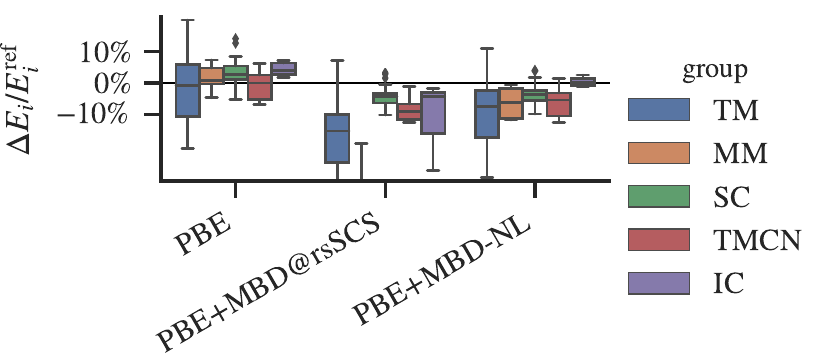}};
\end{tikzpicture}
\caption{\textbf{Distributions of relative errors in binding and lattice energies.}
The boxes show quartiles of the distributions, the whiskers extend further up to 1.5-fold the inter-quartile range, and the individual points denote outliers.
(\textbf{a}) S66$\times$8 set of organic dimers.
The color scale encodes the distance between the centers of mass of the monomers, divided by the equilibrium distance.
(\textbf{b}) Set of 64 hard solids \citep{ZhangNJP18}.
The color encodes the class of a solid: transition metals (TM), main-group metals (MM), semiconductors (SC), transition-metal carbides and nitrides (TMCN), and ionic crystals (IC).
The left-right order of the systems in the plot corresponds to the top-bottom order in the legend.
}\label{fig:errors}
\end{figure}

Next, we briefly describe several benchmark tests of MBD-NL (see \SMref\ for a more detailed description of the calculations and for additional results).
On a set of small organic dimers (S66, \citein{RezacJCTC11}), MBD-NL performs nearly identically to MBD@rsSCS (Figure~\ref{fig:errors}a), which is already excellent for a DFT+vdW approach.
In contrast, the errors in lattice energies of 64 hard solids \citep{ZhangNJP18} are reduced drastically when MBD@rsSCS is replaced with MBD-NL (Figure~\ref{fig:errors}b).
This improvement comes mainly from the errors for metals and ionic solids, which MBD@rsSCS overbinds substantially, whereas plain PBE performs reasonably well, and MBD-NL retains this good performance.
MBD-NL still somewhat overbinds the metals compared to PBE, as could be expected, because bare PBE does not underbind the metals despite the missing (non-jellium) long-range vdW interactions.
Ionic solids are underbound by 4\% with PBE, which is reduced nearly to zero when the missing nonlocal correlation is added by MBD-NL, whereas MBD@rsSCS overbinds some of them substantially.
The performance of MBD-NL on semiconductors is similar to MBD@rsSCS\@.
On a set of organic molecular crystals (X23, \citein{ReillyJCP13}), MBD-NL performs nearly identically to MBD@rsSCS, with a similar tendency to underbind (2\%) as MBD@rsSCS has to overbind (3\%).
On a set of supramolecular complexes (S12L, \citein{GrimmeCEJ12}), the accuracy of MBD-NL is reduced compared to MBD@rsSCS, from 5\% to 9\% in terms of the mean absolute relative error (MARE), but the accuracy of the two methods is equal with the PBE0 functional, with MBD-NL having a smaller mean relative error compared to MBD@rsSCS\@.

Compounds with small or zero electronic gap pose the hardest problem for DFT+vdW approaches, because such systems require in principle long-range coupling of delocalized electronic fluctuations.
Despite that, MBD-NL reaches the accuracy of established effective models for hybrid interfaces of metallic surfaces and organic molecules, such as the MBD@rsSCS[surf] method \citep{RuizPRB16}, with a difference in the binding energy between the two methods below 10\% for a benzene molecule on a silver (111) surface.
This is only possible because the long-wavelength electronic fluctuations in the metal have no correlation counterpart in the adsorbed molecule, so a fully delocalized treatment of the fluctuations is not necessary in this case.

In contrast, the delocalized fluctuations cannot be effectively neglected in layered vdW materials with small band gaps, such as the transition-metal dichalcogenides (TMDCs), which comprise 23 of the benchmark set of 26 layered materials (here dubbed ``26'', \citein{BjorkmanPRB12}).
MBD@rsSCS and VV10 overbind the ``26'' set by 10\% and 52\%, respectively, indicating that both models overpolarize these small-gap layered compounds.
In contrast, the nonlocal part of the polarizability from low-gradient density regions is removed in MBD-NL, resulting in its underbinding of the ``26'' set by 21\% (the accuracy of the reference calculations is 10\%--20\%).
Of the three methods, the three non-TMDC layered materials in the ``26'' set (graphite, BN, PbO) are described most accurately by MBD-NL (MARE of 7\%, compared to 27\% for MBD@rsSCS and 53\% for VV10).


Before concluding, we discuss some open questions regarding MBD-NL\@.
First, the VV polarizability functional is semi-empirical and it can be improved by including nonlocal density information, for example by developing a meta-GGA polarizability functional.
Such an extension could improve the overall accuracy significantly, but requires nontrivial advances in the general theory of polarizability functionals.
Another possibility would be to normalize the vdW parameters not only to free atoms, but also to ions \citep{GouldJCTC16a}, which is comparably more straightforward.
Second, although MBD-NL can effectively treat hybrid interfaces between organic and metallic compounds, it does not capture the truly nonlocal electronic fluctuations that can be found in conductors \citep{DobsonIJQC14}.
Incorporating such a mechanism would not only enable MBD-NL to treat long-range interactions between fully metallic bodies, but also increase its accuracy for interacting systems of small-gap compounds, such as TMDCs.
How to do this in practice is at the moment unclear, and we see this as the largest remaining theoretical gap in the general understanding of vdW interactions in materials.
Third, MBD-NL uses an empirical range-separating function parameterized for a given XC functional.
XC functionals differ substantially in their mid-range behavior (unlike the PBE and PBE0 functionals used here), and developing seamless range-separation approaches that couple semilocal XC functionals and vdW methods in a universal and transferable way remains an open challenge \citep{HermannJCTC18}.


In conclusion, we have developed a vdW model that unifies atomic many-body methods and nonlocal vdW functionals.
By normalizing to free atoms and jellium, we have retained the accuracy of best DFT+vdW approaches while extending applicability to ionic and metallic compounds, and hybrid metal-organic interfaces.
Our approach enables efficient, accurate, and consistent modeling of many-body vdW interactions in a substantially broader range of systems than previously possible.

\begingroup
\setlength\bibsep{0pt}
\footnotesize
\bibliography{refs-zotero,refs}

\newcommand{\noopsort}[1]{}
\begin{thebibliography}{51}
\providecommand{\natexlab}[1]{#1}
\providecommand{\url}[1]{\texttt{#1}}
\expandafter\ifx\csname urlstyle\endcsname\relax
  \providecommand{\doi}[1]{doi: #1}\else
  \providecommand{\doi}{doi: \begingroup \urlstyle{rm}\Url}\fi

\bibitem[Adamo and Barone(1999)]{AdamoJCP99}
C.~Adamo and V.~Barone.
\newblock Toward reliable density functional methods without adjustable
  parameters: {{The PBE0}} model.
\newblock \emph{J. Chem. Phys.}, 110:\penalty0 6158--6170, 1999.
\newblock \doi{10.1063/1.478522}.

\bibitem[Ambrosetti et~al.(2014{\natexlab{a}})Ambrosetti, Alf{\`e}, DiStasio,
  and Tkatchenko]{AmbrosettiJPCL14}
A.~Ambrosetti, D.~Alf{\`e}, R.~A. DiStasio, Jr., and A.~Tkatchenko.
\newblock Hard {{Numbers}} for {{Large Molecules}}: {{Toward Exact Energetics}}
  for {{Supramolecular Systems}}.
\newblock \emph{J. Phys. Chem. Lett.}, 5:\penalty0 849--855,
  2014{\natexlab{a}}.
\newblock \doi{10.1021/jz402663k}.

\bibitem[Ambrosetti et~al.(2014{\natexlab{b}})Ambrosetti, Reilly, DiStasio, and
  Tkatchenko]{AmbrosettiJCP14}
A.~Ambrosetti, A.~M. Reilly, R.~A. DiStasio, Jr., and A.~Tkatchenko.
\newblock Long-range correlation energy calculated from coupled atomic response
  functions.
\newblock \emph{J. Chem. Phys.}, 140:\penalty0 18A508, 2014{\natexlab{b}}.
\newblock \doi{10.1063/1.4865104}.

\bibitem[Becke and Edgecombe(1990)]{BeckeJCP90}
A.~D. Becke and K.~E. Edgecombe.
\newblock A simple measure of electron localization in atomic and molecular
  systems.
\newblock \emph{J. Chem. Phys.}, 92:\penalty0 5397--5403, 1990.
\newblock \doi{10.1063/1.458517}.

\bibitem[Becke and Johnson(2007)]{BeckeJCP07}
A.~D. Becke and E.~R. Johnson.
\newblock Exchange-hole dipole moment and the dispersion interaction revisited.
\newblock \emph{J. Chem. Phys.}, 127:\penalty0 154108, 2007.
\newblock \doi{10.1063/1.2795701}.

\bibitem[Bj{\"o}rkman(2012)]{BjorkmanPRB12}
T.~Bj{\"o}rkman.
\newblock Van der {{Waals}} density functional for solids.
\newblock \emph{Phys. Rev. B}, 86:\penalty0 165109, 2012.
\newblock \doi{10.1103/PhysRevB.86.165109}.

\bibitem[Bj{\"o}rkman et~al.(2012)Bj{\"o}rkman, Gulans, Krasheninnikov, and
  Nieminen]{BjorkmanPRL12}
T.~Bj{\"o}rkman, A.~Gulans, A.~V. Krasheninnikov, and R.~M. Nieminen.
\newblock Van der {{Waals Bonding}} in {{Layered Compounds}} from {{Advanced
  Density}}-{{Functional First}}-{{Principles Calculations}}.
\newblock \emph{Phys. Rev. Lett.}, 108:\penalty0 235502, 2012.
\newblock \doi{10.1103/PhysRevLett.108.235502}.

\bibitem[Blum et~al.(2009)Blum, Gehrke, Hanke, Havu, Havu, Ren, Reuter, and
  Scheffler]{BlumCPC09}
V.~Blum, R.~Gehrke, F.~Hanke, P.~Havu, V.~Havu, X.~Ren, K.~Reuter, and
  M.~Scheffler.
\newblock Ab initio molecular simulations with numeric atom-centered orbitals.
\newblock \emph{Comput. Phys. Commun.}, 180:\penalty0 2175--2196, 2009.
\newblock \doi{10.1016/j.cpc.2009.06.022}.

\bibitem[Caldeweyher et~al.(2019)Caldeweyher, Ehlert, Hansen, Neugebauer,
  Spicher, Bannwarth, and Grimme]{CaldeweyherJCP19}
E.~Caldeweyher, S.~Ehlert, A.~Hansen, H.~Neugebauer, S.~Spicher, C.~Bannwarth,
  and S.~Grimme.
\newblock A generally applicable atomic-charge dependent {{London}} dispersion
  correction.
\newblock \emph{J. Chem. Phys.}, 150:\penalty0 154122, 2019.
\newblock \doi{10.1063/1.5090222}.

\bibitem[Dion et~al.(2004)Dion, Rydberg, Schr{\"o}der, Langreth, and
  Lundqvist]{DionPRL04}
M.~Dion, H.~Rydberg, E.~Schr{\"o}der, D.~C. Langreth, and B.~I. Lundqvist.
\newblock Van der {{Waals}} density functional for general geometries.
\newblock \emph{Phys. Rev. Lett.}, 92:\penalty0 246401, 2004.
\newblock \doi{10.1103/PhysRevLett.92.246401}.

\bibitem[Dobson(2014)]{DobsonIJQC14}
J.~F. Dobson.
\newblock Beyond pairwise additivity in {{London}} dispersion interactions.
\newblock \emph{Int. J. Quantum Chem.}, 114:\penalty0 1157--1161, 2014.
\newblock \doi{10.1002/qua.24635}.

\bibitem[Fedorov et~al.(2018)Fedorov, Sadhukhan, St{\"o}hr, and
  Tkatchenko]{FedorovPRL18}
D.~V. Fedorov, M.~Sadhukhan, M.~St{\"o}hr, and A.~Tkatchenko.
\newblock Quantum-{{Mechanical Relation}} between {{Atomic Dipole
  Polarizability}} and the van der {{Waals Radius}}.
\newblock \emph{Phys. Rev. Lett.}, 121:\penalty0 183401, 2018.
\newblock \doi{10.1103/PhysRevLett.121.183401}.

\bibitem[Ferri et~al.(2015)Ferri, DiStasio, Ambrosetti, Car, and
  Tkatchenko]{FerriPRL15}
N.~Ferri, R.~A. DiStasio, Jr., A.~Ambrosetti, R.~Car, and A.~Tkatchenko.
\newblock Electronic {{Properties}} of {{Molecules}} and {{Surfaces}} with a
  {{Self}}-{{Consistent Interatomic}} van der {{Waals Density Functional}}.
\newblock \emph{Phys. Rev. Lett.}, 114:\penalty0 176802, 2015.
\newblock \doi{10.1103/PhysRevLett.114.176802}.

\bibitem[Gould et~al.(2016)Gould, Leb{\`e}gue, {\'A}ngy{\'a}n, and Bu{\v
  c}ko]{GouldJCTC16a}
T.~Gould, S.~Leb{\`e}gue, J.~G. {\'A}ngy{\'a}n, and T.~Bu{\v c}ko.
\newblock A {{Fractionally Ionic Approach}} to {{Polarizability}} and van der
  {{Waals Many}}-{{Body Dispersion Calculations}}.
\newblock \emph{J. Chem. Theory Comput.}, 12:\penalty0 5920--5930, 2016.
\newblock \doi{10.1021/acs.jctc.6b00925}.

\bibitem[Grimme(2012)]{GrimmeCEJ12}
S.~Grimme.
\newblock Supramolecular {{Binding Thermodynamics}} by
  {{Dispersion}}-{{Corrected Density Functional Theory}}.
\newblock \emph{Chem. Eur. J.}, 18:\penalty0 9955--9964, 2012.
\newblock \doi{10.1002/chem.201200497}.

\bibitem[Grimme et~al.(2010)Grimme, Antony, Ehrlich, and Krieg]{GrimmeJCP10}
S.~Grimme, J.~Antony, S.~Ehrlich, and H.~Krieg.
\newblock A consistent and accurate ab initio parametrization of density
  functional dispersion correction ({{DFT}}-{{D}}) for the 94 elements
  {{H}}-{{Pu}}.
\newblock \emph{J. Chem. Phys.}, 132:\penalty0 154104, 2010.
\newblock \doi{10.1063/1.3382344}.

\bibitem[Grimme et~al.(2016)Grimme, Hansen, Brandenburg, and
  Bannwarth]{GrimmeCR16}
S.~Grimme, A.~Hansen, J.~G. Brandenburg, and C.~Bannwarth.
\newblock Dispersion-{{Corrected Mean}}-{{Field Electronic Structure Methods}}.
\newblock \emph{Chem. Rev.}, 116:\penalty0 5105--5154, 2016.
\newblock \doi{10.1021/acs.chemrev.5b00533}.

\bibitem[Gutle et~al.(1999)Gutle, Savin, Krieger, and Chen]{GutleIJQC99}
C.~Gutle, A.~Savin, J.~B. Krieger, and J.~Chen.
\newblock Correlation energy contributions from low-lying states to density
  functionals based on an electron gas with a gap.
\newblock \emph{Int. J. Quantum Chem.}, 75:\penalty0 885--888, 1999.
\newblock \doi{10.1002/(SICI)1097-461X(1999)75:4/5<885::AID-QUA53>3.0.CO;2-F}.

\bibitem[Hermann(2019{\natexlab{a}})]{Code}
J.~Hermann.
\newblock 2019{\natexlab{a}}.
\newblock \doi{10.6084/m9.figshare.9943361.v2}.
\newblock Code as git repository.

\bibitem[Hermann(2019{\natexlab{b}})]{Data}
J.~Hermann.
\newblock 2019{\natexlab{b}}.
\newblock \doi{10.6084/m9.figshare.9943301.v1}.
\newblock Data in HDF5 format.

\bibitem[Hermann(2019{\natexlab{c}})]{Libmbd}
J.~Hermann.
\newblock Libmbd.
\newblock 2019{\natexlab{c}}.
\newblock \doi{10.5281/zenodo.3474093}.
\newblock Code as git repository.

\bibitem[Hermann and Tkatchenko(2018)]{HermannJCTC18}
J.~Hermann and A.~Tkatchenko.
\newblock Electronic exchange and correlation in van der {{Waals}} systems:
  {{Balancing}} semilocal and nonlocal energy contributions.
\newblock \emph{J. Chem. Theory Comput.}, 14:\penalty0 1361--1369, 2018.
\newblock \doi{10.1021/acs.jctc.7b01172}.

\bibitem[Hermann et~al.(2017)Hermann, DiStasio, and Tkatchenko]{HermannCR17}
J.~Hermann, R.~A. DiStasio, Jr., and A.~Tkatchenko.
\newblock First-principles models for van der {{Waals}} interactions in
  molecules and materials: Concepts, theory, and applications.
\newblock \emph{Chem. Rev.}, 117:\penalty0 4714--4758, 2017.
\newblock \doi{10.1021/acs.chemrev.6b00446}.

\bibitem[Hirshfeld(1977)]{HirshfeldTCA77}
F.~L. Hirshfeld.
\newblock Bonded-atom fragments for describing molecular charge densities.
\newblock \emph{Theor. Chim. Acta}, 44:\penalty0 129--138, 1977.
\newblock \doi{10.1007/BF00549096}.

\bibitem[Jaffe(2005)]{JaffePRD05}
R.~L. Jaffe.
\newblock Casimir effect and the quantum vacuum.
\newblock \emph{Phys. Rev. D}, 72:\penalty0 021301(R), 2005.
\newblock \doi{10.1103/PhysRevD.72.021301}.

\bibitem[Kim et~al.(2020)Kim, Kim, Gould, Lee, Leb{\`e}gue, and Kim]{KimJACS20}
M.~Kim, W.~J. Kim, T.~Gould, E.~K. Lee, S.~Leb{\`e}gue, and H.~Kim.
\newblock {{uMBD}}: {{A Materials}}-{{Ready Dispersion Correction That
  Uniformly Treats Metallic}}, {{Ionic}}, and van der {{Waals Bonding}}.
\newblock \emph{J. Am. Chem. Soc.}, 142:\penalty0 2346--2354, 2020.
\newblock \doi{10.1021/jacs.9b11589}.

\bibitem[Klime{\v s} and Michaelides(2012)]{KlimesJCP12}
J.~Klime{\v s} and A.~Michaelides.
\newblock Perspective: {{Advances}} and challenges in treating van der
  {{Waals}} dispersion forces in density functional theory.
\newblock \emph{J. Chem. Phys.}, 137:\penalty0 120901, 2012.
\newblock \doi{10.1063/1.4754130}.

\bibitem[K{\"u}mmel and Perdew(2003)]{KummelMP03}
S.~K{\"u}mmel and J.~P. Perdew.
\newblock Two avenues to self-interaction correction within
  {{Kohn}}\textendash{{Sham}} theory: Unitary invariance is the shortcut.
\newblock \emph{Mol. Phys.}, 101:\penalty0 1363--1368, 2003.
\newblock \doi{10.1080/0026897031000094506}.

\bibitem[London(1930)]{LondonZP30}
F.~London.
\newblock {Zur Theorie und Systematik der Molekularkr{\"a}fte [On theory and
  classification of molecular forces]}.
\newblock \emph{Z. Physik}, 63:\penalty0 245--279, 1930.
\newblock \doi{10.1007/BF01421741}.

\bibitem[Lu et~al.(2009)Lu, Li, Rocca, and Galli]{LuPRL09}
D.~Lu, Y.~Li, D.~Rocca, and G.~Galli.
\newblock Ab initio {{Calculation}} of van der {{Waals Bonded Molecular
  Crystals}}.
\newblock \emph{Phys. Rev. Lett.}, 102:\penalty0 206411, 2009.
\newblock \doi{10.1103/PhysRevLett.102.206411}.

\bibitem[Lucas(1967)]{LucasP67}
A.~Lucas.
\newblock Collective contributions to the long-range dipolar interaction in
  rare-gas crystals.
\newblock \emph{Physica}, 35:\penalty0 353--368, 1967.
\newblock \doi{10.1016/0031-8914(67)90184-X}.

\bibitem[McLachlan(1963)]{McLachlanPRSLA63}
A.~D. McLachlan.
\newblock Retarded {{Dispersion Forces Between Molecules}}.
\newblock \emph{Proc. Royal Soc. Lond. A}, 271:\penalty0 387--401, 1963.
\newblock \doi{10.1098/rspa.1963.0025}.

\bibitem[Perdew et~al.(1996)Perdew, Burke, and Ernzerhof]{PerdewPRL96}
J.~P. Perdew, K.~Burke, and M.~Ernzerhof.
\newblock Generalized {{Gradient Approximation Made Simple}}.
\newblock \emph{Phys. Rev. Lett.}, 77:\penalty0 3865--3868, 1996.
\newblock \doi{10.1103/PhysRevLett.77.3865}.

\bibitem[Reilly and Tkatchenko(2013)]{ReillyJCP13}
A.~M. Reilly and A.~Tkatchenko.
\newblock Understanding the role of vibrations, exact exchange, and many-body
  van der {{Waals}} interactions in the cohesive properties of molecular
  crystals.
\newblock \emph{J. Chem. Phys.}, 139:\penalty0 024705, 2013.
\newblock \doi{10.1063/1.4812819}.

\bibitem[{\v R}ez{\'a}{\v c} et~al.(2011){\v R}ez{\'a}{\v c}, Riley, and
  Hobza]{RezacJCTC11}
J.~{\v R}ez{\'a}{\v c}, K.~E. Riley, and P.~Hobza.
\newblock S66: {{A Well}}-balanced {{Database}} of {{Benchmark Interaction
  Energies Relevant}} to {{Biomolecular Structures}}.
\newblock \emph{J. Chem. Theory Comput.}, 7:\penalty0 2427--2438, 2011.
\newblock \doi{10.1021/ct2002946}.

\bibitem[Ruiz et~al.(2012)Ruiz, Liu, Zojer, Scheffler, and
  Tkatchenko]{RuizPRL12}
V.~G. Ruiz, W.~Liu, E.~Zojer, M.~Scheffler, and A.~Tkatchenko.
\newblock Density-{{Functional Theory}} with {{Screened}} van der {{Waals
  Interactions}} for the {{Modeling}} of {{Hybrid Inorganic}}-{{Organic
  Systems}}.
\newblock \emph{Phys. Rev. Lett.}, 108:\penalty0 146103, 2012.
\newblock \doi{10.1103/PhysRevLett.108.146103}.

\bibitem[Ruiz et~al.(2016)Ruiz, Liu, and Tkatchenko]{RuizPRB16}
V.~G. Ruiz, W.~Liu, and A.~Tkatchenko.
\newblock Density-functional theory with screened van der {{Waals}}
  interactions applied to atomic and molecular adsorbates on close-packed and
  non-close-packed surfaces.
\newblock \emph{Phys. Rev. B}, 93:\penalty0 035118, 2016.
\newblock \doi{10.1103/PhysRevB.93.035118}.

\bibitem[Sato and Nakai(2009)]{SatoJCP09}
T.~Sato and H.~Nakai.
\newblock Density functional method including weak interactions: {{Dispersion}}
  coefficients based on the local response approximation.
\newblock \emph{J. Chem. Phys.}, 131:\penalty0 224104, 2009.
\newblock \doi{10.1063/1.3269802}.

\bibitem[Sato and Nakai(2010)]{SatoJCP10}
T.~Sato and H.~Nakai.
\newblock Local response dispersion method. {{II}}. {{Generalized}} multicenter
  interactions.
\newblock \emph{J. Chem. Phys.}, 133:\penalty0 194101, 2010.
\newblock \doi{10.1063/1.3503040}.

\bibitem[Silvestrelli(2008)]{SilvestrelliPRL08}
P.~L. Silvestrelli.
\newblock Van der {{Waals Interactions}} in {{DFT Made Easy}} by {{Wannier
  Functions}}.
\newblock \emph{Phys. Rev. Lett.}, 100:\penalty0 053002, 2008.
\newblock \doi{10.1103/PhysRevLett.100.053002}.

\bibitem[Silvestrelli(2013)]{SilvestrelliJCP13}
P.~L. Silvestrelli.
\newblock Van der {{Waals}} interactions in density functional theory by
  combining the quantum harmonic oscillator-model with localized {{Wannier}}
  functions.
\newblock \emph{J. Chem. Phys.}, 139:\penalty0 054106, 2013.
\newblock \doi{10.1063/1.4816964}.

\bibitem[Sun et~al.(2013)Sun, Xiao, Fang, Haunschild, Hao, Ruzsinszky, Csonka,
  Scuseria, and Perdew]{SunPRL13}
J.~Sun, B.~Xiao, Y.~Fang, R.~Haunschild, P.~Hao, A.~Ruzsinszky, G.~I. Csonka,
  G.~E. Scuseria, and J.~P. Perdew.
\newblock Density {{Functionals}} that {{Recognize Covalent}}, {{Metallic}},
  and {{Weak Bonds}}.
\newblock \emph{Phys. Rev. Lett.}, 111:\penalty0 106401, 2013.
\newblock \doi{10.1103/PhysRevLett.111.106401}.

\bibitem[Sun et~al.(2015)Sun, Ruzsinszky, and Perdew]{SunPRL15}
J.~Sun, A.~Ruzsinszky, and J.~P. Perdew.
\newblock Strongly {{Constrained}} and {{Appropriately Normed Semilocal Density
  Functional}}.
\newblock \emph{Phys. Rev. Lett.}, 115:\penalty0 036402, 2015.
\newblock \doi{10.1103/PhysRevLett.115.036402}.

\bibitem[Tkatchenko and Scheffler(2009)]{TkatchenkoPRL09}
A.~Tkatchenko and M.~Scheffler.
\newblock Accurate {{Molecular Van Der Waals Interactions}} from
  {{Ground}}-{{State Electron Density}} and {{Free}}-{{Atom Reference Data}}.
\newblock \emph{Phys. Rev. Lett.}, 102:\penalty0 073005, 2009.
\newblock \doi{10.1103/PhysRevLett.102.073005}.

\bibitem[Tkatchenko et~al.(2012)Tkatchenko, DiStasio, Car, and
  Scheffler]{TkatchenkoPRL12}
A.~Tkatchenko, R.~A. DiStasio, Jr., R.~Car, and M.~Scheffler.
\newblock Accurate and {{Efficient Method}} for {{Many}}-{{Body}} van der
  {{Waals Interactions}}.
\newblock \emph{Phys. Rev. Lett.}, 108:\penalty0 236402, 2012.
\newblock \doi{10.1103/PhysRevLett.108.236402}.

\bibitem[Tkatchenko et~al.(2013)Tkatchenko, Ambrosetti, and
  DiStasio]{TkatchenkoJCP13}
A.~Tkatchenko, A.~Ambrosetti, and R.~A. DiStasio, Jr.
\newblock Interatomic methods for the dispersion energy derived from the
  adiabatic connection fluctuation-dissipation theorem.
\newblock \emph{J. Chem. Phys.}, 138:\penalty0 74106, 2013.
\newblock \doi{10.1063/1.4789814}.

\bibitem[Vydrov and Van~Voorhis(2009)]{VydrovPRL09}
O.~A. Vydrov and T.~Van~Voorhis.
\newblock Nonlocal van der {{Waals Density Functional Made Simple}}.
\newblock \emph{Phys. Rev. Lett.}, 103:\penalty0 063004, 2009.
\newblock \doi{10.1103/PhysRevLett.103.063004}.

\bibitem[Vydrov and Van~Voorhis(2010{\natexlab{a}})]{VydrovJCP10a}
O.~A. Vydrov and T.~Van~Voorhis.
\newblock Nonlocal van der {{Waals}} density functional: {{The}} simpler the
  better.
\newblock \emph{J. Chem. Phys.}, 133:\penalty0 244103, 2010{\natexlab{a}}.
\newblock \doi{10.1063/1.3521275}.

\bibitem[Vydrov and Van~Voorhis(2010{\natexlab{b}})]{VydrovPRA10}
O.~A. Vydrov and T.~Van~Voorhis.
\newblock Dispersion interactions from a local polarizability model.
\newblock \emph{Phys. Rev. A}, 81:\penalty0 062708, 2010{\natexlab{b}}.
\newblock \doi{10.1103/PhysRevA.81.062708}.

\bibitem[Yang et~al.(2014)Yang, Hu, Usvyat, Matthews, Sch{\"u}tz, and
  Chan]{YangS14}
J.~Yang, W.~Hu, D.~Usvyat, D.~Matthews, M.~Sch{\"u}tz, and G.~K.-L. Chan.
\newblock Ab initio determination of the crystalline benzene lattice energy to
  sub-kilojoule/mole accuracy.
\newblock \emph{Science}, 345:\penalty0 640--643, 2014.
\newblock \doi{10.1126/science.1254419}.

\bibitem[Zhang et~al.(2018)Zhang, Reilly, Tkatchenko, and
  Scheffler]{ZhangNJP18}
G.-X. Zhang, A.~M. Reilly, A.~Tkatchenko, and M.~Scheffler.
\newblock Performance of various density-functional approximations for cohesive
  properties of 64 bulk solids.
\newblock \emph{New J. Phys.}, 20:\penalty0 063020, 2018.
\newblock \doi{10.1088/1367-2630/aac7f0}.

\end{thebibliography}
\endgroup

\section*{Appendix}


All computational resources for the manuscript can be found in a Git repository \citep{Code} and related data files \citep{Data}.
This includes scripts used to generate input files, to run the calculations, process and analyze data, and generate figures.
The file organization is described in the \verb+README.md+ file in the repository.

All DFT calculations were done with FHI-aims \citep{BlumCPC09}, which uses atom-centered basis sets with numerical radial parts.
We used the \verb+tight+ default basis set and grid settings, which ensure numerical convergence to 0.1\,kcal/mol in binding energies for the van der Waals (vdW) systems studied here.
MBD calculations were performed with the help of the Libmbd library \citep{Libmbd}, which is integrated into FHI-aims, and MBD-NL calculations can be performed directly in FHI-aims with a current development version.
Our current implementation does not include the functional derivative $\delta\alpha^\text{VV'}[n]/\delta n$, and as such MBD-NL is evaluated on the self-consistent PBE densities in this work, while the implementation of the derivative is a work in progress.
Importantly, the electron density change induced by vdW interactions has been found to have only a negligible effect on the interaction energies and nuclear forces \citep{FerriPRL15}.
The PBE, PBE0, and VV10 energies for the S66, X23, and S12L sets were taken from \citep{HermannJCTC18}, which used the same numerical settings as this work.
For molecular crystals, $k$-point grids with density of at least 0.8\,\AA\ in reciprocal space were used for all DFT and MBD calculations.
For hard solids, we have used the $k$-point density from \citep{ZhangNJP18}.
All molecular and crystal geometries were taken directly from the respective benchmark sets without any relaxation.

\begin{table}[t]
\centering
\caption{\textbf{Errors in interaction energies on vdW benchmark data sets.}}\label{tab:performance}
\small
\begin{tabular}{%
  ll
  *{2}{S[
    table-format=2.1,
    table-align-text-post=false,
    table-space-text-post=\%,
  ]}
  S[
    table-format=2.1,
    table-align-text-post=false,
  ]
  S[
    table-format=-2.1,
    table-align-text-pre=false,
    table-align-text-post=false,
    table-space-text-pre=k,
  ]
}
\toprule
Method & & {S66} & {X23} & {S12L} & {``26''$^a$} \\
\midrule
PBE        & MRE$^b$  & 57\%   & 60\%   & 82\%   & 105\%      \\
\midrule[0.1pt]
+MBD@rsSCS & MARE$^c$ & 8.4\%  & 6.4\%  & 5.3\%  & 14\%$^d$ \\
           & MRE      & -3.1\% & -3.4\% & -1.4\% & -10\%$^d$    \\
\midrule[0.1pt]
+MBD-NL    & MARE     & 9.3\%  & 6.2 \% & 9.2\%  & 21\%       \\
           & MRE      & -0.1\% & 1.9\%  & 6.4\%  & 21\%       \\
\midrule[0.1pt]
+VV10      & MARE     & 9.9\%  & 15\%   & 15\%   & 52\%$^e$ \\
           & MRE      & -6.1\% & -15\%  & -15\%  & -52\%$^e$    \\
\midrule
PBE0       & MRE  & 56\%   & 58\%   & 75\%   & \\
\midrule[0.1pt]
+MBD@rsSCS & MARE & 7.6\%  & 5.4\%  & 6.5\%  & \\
           & MRE  & -1.1\% & -1.7\% & -4.4\% & \\
\midrule[0.1pt]
+MBD-NL    & MARE & 8.5 \% & 5.7\%  & 6.8\%  & \\
           & MRE  & 0.0\%  & 3.0\%  & 1.7\%  & \\
\midrule[0.1pt]
+VV10      & MARE & 8.3\%  & 15\%   & 20\%   & \\
           & MRE  & -5.3\% & -15\%  & -20\%  & \\
\bottomrule
\end{tabular}

\begin{minipage}{.96\linewidth}
\footnotesize%
$^a$Data set of interlayer binding energies of 26 layered materials with RPA benchmark energies by \citet{BjorkmanPRL12}.
$^b$Mean relative error. Negative numbers indicate overbinding.
$^c$Mean absolute relative error.
$^d$The eigenvalue rescaling for MBD by \citet{GouldJCTC16a} must be used, otherwise the MBD Hamiltonian has nonnegative eigenvalues for only 6 compounds (graphite, BN, PbO, and 3 TMDCs).
$^e$Results as given by \citet{BjorkmanPRB12} for the original PW86r+VV10 combination.
\end{minipage}
\end{table}

Table~\ref{tab:performance} reports the performance of MBD-NL, MBD@rsSCS, and VV10, in combination with the PBE and PBE0 functionals, on the set of organic molecular crystals (X23, \citein{ReillyJCP13}), a set of supramolecular complexes (S12L, \citein{GrimmeCEJ12}), and a set of 26 layered materials (dubbed ``26'', \citein{BjorkmanPRB12}).
Of the standard vdW data sets, S12L is the only one where MBD-NL achieves a different performance with the PBE and PBE0 functionals.
This results mostly from PBE binding the $\mathrm\pi$--$\mathrm\pi$ complexes somewhat more than PBE0.
The proper balance between semilocal DFT and long-range vdW models in the case of large $\mathrm\pi$--$\mathrm\pi$ complexes is unclear \citep{HermannJCTC18}.
On the ``26'' set, the MBD@rsSCS Hamiltonian has negative eigenvalues for 20 of the 26 compounds.
To obtain finite energies nevertheless, we use the eigenvalue rescaling as proposed by \citet{GouldJCTC16a}.

\begin{figure}[t!]
\centering
\includegraphics{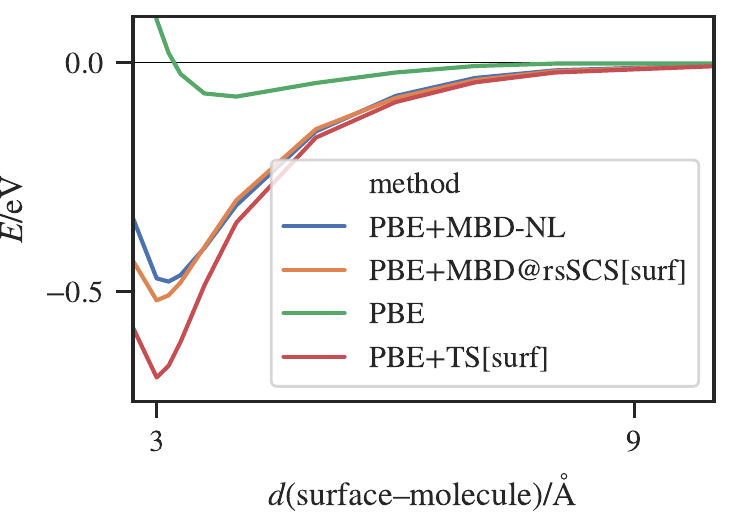}
\caption{\textbf{Binding energy of a single benzene molecule on a (111) silver surface.}
}\label{fig:silver-benzene}
\end{figure}

Figure~\ref{fig:silver-benzene} compares the binding energy curve of a hybrid organic/inorganic interface as calculated by PBE-NL and surface variants of the MBD@rsSCS and TS methods by \citet{RuizPRB16}.

\begin{table}[t!]
\centering
\caption{\textbf{Timing of DFT and MBD calculations}}\label{tab:timing}
\small
\begin{tabular}{l*2{S[table-format=5.1]}}
\toprule
& \multicolumn2c{CPU time$^a$ [s]} \\
Calculation & {MoS$_2$$^b$} & {4a @ S12L$^c$} \\
\midrule
Complete KS-DFT        & 3700 & 39000 \\
Single KS-DFT cycle    & 250  & 1900 \\
Kinetic energy density & 120  & 31 \\
MBD energy             & 0.6  & 1.2 \\
\bottomrule
\end{tabular}

\begin{minipage}{.7\linewidth}
\footnotesize
$^a$Total single-core CPU time on an Intel Xeon Gold 6148 processor (Skylake).
$^b$6 atoms in a unit cell, 200 $k$-points.
$^c$148 atoms.
\end{minipage}
\end{table}

Table~\ref{tab:timing} presents timings of illustrative DFT+MBD calculations and their individual components for a simple inorganic material and a large organic complex.
In both cases, the evaluation of the MBD energy is only a small fraction of the cost of the DFT calculation, even when the evaluation of the kinetic energy density needed for parametrization of MBD-NL is included in the cost of the MBD calculation.

\begin{figure}[t!]
\centering
\includegraphics{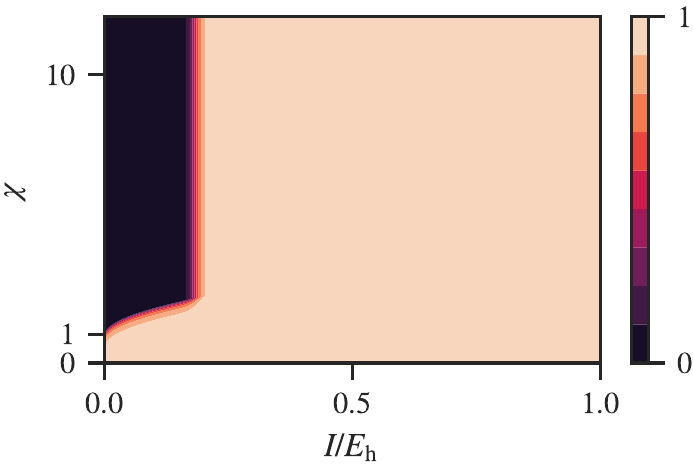}
\caption{\textbf{Smooth cutoff function for the low-gradient contributions to the polarizability.}
}\label{fig:cutoff-func}
\end{figure}

The function $g(I,\chi)$ from eq.\ (6) of the main text is visualized in Figure~\ref{fig:cutoff-func}.

\end{document}